\documentclass{ws-procs9x6}

\def\gtwid{\mathrel{\raise.3ex\hbox{$>$\kern-.75em\lower1ex\hbox{$\sim$}}}}
\def\ltwid{\mathrel{\raise.3ex\hbox{$<$\kern-.75em\lower1ex\hbox{$\sim$}}}}

\begin{document}

\title{DARK MATTER AXIONS}

\author{P. SIKIVIE}

\address{Department of Physics, University of Florida,\\
Gainesville, FL 32611, USA\\
E-mail: sikivie@phys.ufl.edu}

\begin{abstract}

The hypothesis of an `invisible' axion was made by Misha Shifman and 
others, approximately thirty years ago.  It has turned out to be an 
unusually fruitful idea, crossing boundaries between particle physics, 
astrophysics and cosmology.  An axion with mass of order $10^{-5}$ eV
(with large uncertainties) is one of the leading candidates for the 
dark matter of the universe.  It was found recently that dark matter 
axions thermalize and form a Bose-Einstein condensate (BEC).  Because 
they form a BEC, axions differ from ordinary cold dark matter (CDM) 
in the non-linear regime of structure formation and upon entering the 
horizon.  Axion BEC provides a mechanism for the production of net 
overall rotation in dark matter halos, and for the alignment of cosmic 
microwave anisotropy multipoles.  Because there is evidence for these 
phenomena, unexplained with ordinary CDM, an argument can be made that 
the dark matter is axions.

\end{abstract}


\bodymatter

\section{Introduction}

It is a great pleasure and honor to be part of Misha Shifman's 60th 
birthday celebration.  Among Misha's many outstanding contributions
to particle physics is his well-known proposal, in collaboration with 
Arkady Vainshtein and Valentine Zakharov, that the axion may be very 
light and very weakly coupled \cite{Shifman80,Kim80,DFSZ}.  Here is 
the abstract of their paper:
\vskip 0.1in
{\it 
P- and T-invariance violation in quantum chromodynamics due to the 
so-called $\theta$-term is discussed. It is shown that irrespectively 
of how the confinement works there emerge observable P- and T-odd effects. 
The proof is based on the assumption that QCD resolves the U(1) problem, 
i.e., the mass of the singlet pseudoscalar meson does not vanish in the 
chiral limit. We suggest a modification of the axion scheme which restores 
the natural P and T invariance of the theory and cannot be ruled out
experimentally.}

The $\theta$-term mentioned by Shifman et al. is 
\begin{equation}
{\cal L}_\theta = 
~+~{\theta g^2\over 32\pi^2} G^a_{\mu\nu} \tilde G^{a\mu\nu}
\label{ggdual}
\end{equation}
where $G^a_{\mu\nu}$ are the QCD field strengths, $g$ is the QCD
coupling constant and $\theta$ is a parameter.  A $\theta$-term is 
generally present in the action density of the Standard Model of 
elementary particles \cite{'tHooft}.  Its existence raises a puzzle, 
called the Strong CP Problem.  As Shifman et al. explain in their 
paper, the physics of QCD necessarily depends on the value of $\theta$, 
if none of the quark masses vanish, because otherwise QCD wouldn't solve 
the U$_{\rm A}$(1) problem (of explaining why the mass of the singlet 
pseudoscalar meson does not vanish in the chiral limit) and hence 
couldn't be the correct theory of strong interactions.  This is an 
important point.  If it were possible for QCD to be independent of 
$\theta$, the Strong CP Problem wouldn't be so urgent.

One can show that QCD physics depends on the value of $\theta$ only 
through the combination $\bar{\theta} \equiv \theta - \arg \det m_q$ 
where $m_q$ is the quark mass matrix.  If $\bar{\theta} \neq 0$ the 
strong interactions violate P and CP.  Such P and CP violation is 
incompatible with the experimental upper bound on the neutron electic 
dipole moment \cite{ned} unless $|\bar{\theta}| < 10^{-10}$.  In the 
Standard Model, P and CP violation are introduced by letting the 
elements of the quark mass matrix $m_q$ be arbitrary complex numbers 
\cite{KM}.  In that case, $\bar{\theta}$ is of order one.  The Strong 
CP Problem is the problem of explaining why $|\bar{\theta}| < 10^{-10}$
instead.

The Strong CP Problem is solved if the term (\ref{ggdual}) in the 
Standard Model action density is replaced by 
\begin{equation}
{\cal L}_{\rm axion} = 
~-~{1 \over 2}\partial_\mu \varphi \partial^\mu \varphi
+ {g^2\over 32\pi^2}~{\varphi(x) \over f}~G^a_{\mu\nu} \tilde G^{a\mu\nu}
\label{ax}
\end{equation}
where $\varphi(x)$ is a new scalar field, and $f$ is a constant with 
dimension of energy.  In the modified theory,~~ 
$\bar{\theta} = {\varphi(x) \over f} - \arg\det m_q$~~ depends on the
expectation value of $\varphi(x)$.  This field settles to a value that
minimizes the effective potential.  The Strong CP Problem is solved
because the minimum of the QCD effective potential $V(\bar{\theta})$
occurs at $\bar{\theta} = 0$ \cite{VW}.  The $\varphi G \cdot \tilde G$
interaction in Eq.~(\ref{ax}) is not renormalizable.  However, there
is a recipe for constructing renormalizable theories whose low energy
effective action density is of the form of Eq.~(\ref{ax}):  construct
the theory in such a way that it has a U$_{\rm PQ}(1)$ symmetry which 
is a global symmetry of the classical action density, is broken by 
the color anomaly, and is spontaneously broken.  Such a symmetry is
called Peccei-Quinn symmetry after its inventors \cite{PQ}.  Weinberg
and Wilczek \cite{WW} pointed out that a theory with U$_{\rm PQ}(1)$
symmetry has a light pseudo-scalar particle, called the axion.  The
axion field is $\varphi(x)$.  $f$ is of order the expectation value that
breaks U$_{\rm PQ}(1)$, and is called the ``axion decay constant".

The axion mass is given in terms of $f$ by \cite{WW}
\begin{equation}
m \simeq 6~{\rm eV}~{10^6~{\rm GeV}\over f}\, .
\label{ma}
\end{equation}
All axion couplings are inversely proportional to $f$.  The
axion coupling to two photons is:
\begin{equation}
{\cal L}_{a\gamma\gamma} = -g_\gamma {\alpha\over \pi} {\varphi(x)\over f}
\vec E \cdot\vec B~~~\ ,
\label{aEB}
\end{equation}
where $\vec E$ and $\vec B$ are the electric and magnetic fields,   
$\alpha$ is the fine structure constant, and $g_\gamma$ is a
model-dependent coefficient of order one.  It had first been 
thought that $f$ is of order the electroweak scale, in which 
case the axion couplings have strength typical of neutrinos 
and the axion mass is relatively large, in the 10 keV to 10 MeV
range.   Such axions were quickly ruled out by particle physics 
(beam dumps and rare decays) and nuclear physics experiments.  
But Shifman and others showed \cite{Shifman80,Kim80,DFSZ} that 
$f$ can be made arbitrarily large, and hence the axion can be 
made arbitrarily light and weakly coupled.  Such an axion is 
unconstrained by the aforementioned experiments, and was dubbed 
`invisible'.  May an invisible axion really exist?  

The axion has been searched for in many places, and has not 
been found \cite{arev}.  Axion masses larger than about 50 keV 
are ruled out by the aforementioned particle and nuclear physics
experiments.  The next range of axion masses, in decreasing order, 
is ruled out by stellar evolution arguments.  The longevity of red 
giants rules out 200 keV $> m >$ 0.5 eV \cite{Dicus,Raff87} in case 
the axion has negligible coupling to the electron (such an axion is 
called `hadronic'), and 200 keV $> m > 10^{-2}$ eV \cite{Schramm}
in case the axion has a sizable coupling to electrons.  The 
duration of the neutrino pulse from supernova 1987a rules out
2 eV $> m > 3 \cdot 10^{-3}$ eV \cite{1987a}.  Finally, there is
a lower limit, $m \gtwid 10^{-6}$ eV, from cosmology which is 
discussed in the next section.  This leaves open an ``axion window":
$3 \cdot 10^{-3} > m \gtwid 10^{-6}$ eV.  The lower edge of this 
window ($10^{-6}$ eV) is much softer than its upper edge.

\section{Axion production in the early universe}

There are two populations of axions produced in the early universe, 
which we may call 'hot' (or thermal) and cold.  Hot axions are produced 
in thermal processes such as $q + g \rightarrow q + a$ where $q$ is a 
quark and $g$ a gluon, or $\pi + \pi \rightarrow \pi + a$ where $\pi$ 
is a pion \cite{KT,Masso02,axcosm}.  The number density of thermal 
axions today (time $t_0$) is 
\begin{equation}
n_a^{\rm th}(t_0) \simeq {7.5 \over {\rm cm}^3} 
\left({106.75 \over {\cal N}_{\rm D}}\right)^{1 \over 3}
\label{thaxden}
\end{equation}
where ${\cal N}_{\rm D}$ is the effective number of thermal degrees 
of freedom at the time axions decouple from the thermal bath.  The 
Standard Model has ${\cal N}_{\rm D} = 106.75$.  Thermal axions are 
a form of hot dark matter, similar to neutrinos, in the context of 
large scale structure formation.

The cold axions are produced when the potential $V(\bar{\theta})$, and 
hence the axion mass, turns on near the QCD phase transition \cite{ac}.  
The critical time, defined by $m(t_1)t_1 = 1$, is 
$t_1 \simeq 2\cdot 10^{-7}$ sec $(f/ 10^{12} {\rm GeV})^{1 \over 3}$.
Cold axions are the quanta of oscillation of the axion field that result
from the turn on of the axion mass.  The average number density of cold 
axions at time $t_1$ is
\begin{equation}
n_a(t_1)\simeq {1\over 2} m(t_1) \langle \varphi^2(t_1)\rangle \simeq
\pi f^2 {1\over t_1}~~~\ .
\label{nat1}
\end{equation}
In Eq.~(\ref{nat1}), we used the fact that the axion field $\varphi(x)$
is approximately homogeneous on the horizon scale $t_1$, because 
wiggles in $\varphi(x)$ which entered the horizon long before $t_1$ 
have been red-shifted away \cite{Vil}.  We also used the fact that 
the initial departure of $\varphi(x)$ from the CP conserving minimum 
is of order $f$.  The axions of Eq.~(\ref{nat1}) are non-relativistic.  
Assuming that the ratio of the axion number density to the entropy 
density is constant from time $t_1$ till today, one finds 
\cite{ac}
\begin{equation}
\Omega_a \simeq {1 \over 2}
\left({f \over 10^{12}{\rm GeV}}\right)^{7\over 6}
\left({0.7 \over h}\right)^2
\label{oma}
\end{equation}
for the ratio of the axion energy density to the critical density
for closing the universe.  $h$ is the present Hubble rate in units
of 100 km/s.Mpc.  The requirement that axions do not overclose the
universe implies the constraint $m  \gtwid 6 \cdot 10^{-6}$~eV.
For a more detailed discussion of the production and properties 
of dark matter axions, the reader may wish to consult 
refs. \cite{axcosm,Duffy}.

\section{Dark matter caustics}

It has been established from a variety of observational inputs
that approximately 23\% of the energy density of the universe
is ``cold dark matter" (CDM).  The CDM particles must be non-baryonic, 
and cold.  ``Cold" means that their primordial velocity dispersion is 
small enough that it can be set equal to zero for all practical purposes 
when discussing the formation of large scale structure.  The leading 
candidates for the CDM particles are axions, weakly interacting massive 
particles (WIMPs), e.g. the neutralino in supersymmetric extensions of 
the Standard Model, and sterile neutrinos with mass in the keV range.

A central problem in dark matter studies is the question how CDM 
is distributed in the halos of galaxies, and in particular in the 
halo of our Milky Way galaxy.  Indeed, knowledge of this distribution 
is essential for understanding galactic dynamics and for predicting 
signals in direct and indirect searches for dark matter on Earth. 

Galactic halos are thought to be collisionless fluids and must 
therefore be described in 6-dimensional phase space.  A full 
description gives the phase space distribution $f(\vec{r}, \vec{v}; t)$ 
of the dark matter particles in the halo, i.e. their velocity   
($\vec{v}$) distribution at every position $\vec{r}$.  An important 
simplification occurs in the case of {\it cold} dark matter because 
CDM particles lie in phase space on a thin 3-dimensional hypersurface. 
This fact implies that the velocity distribution is everywhere discrete 
\cite{Ips} and that there are surfaces in physical space, called caustics, 
where the density of dark matter is very large.  

Galactic halos have two types of caustics, outer and inner.  The
outer caustics are simple fold ($A_2$) catastrophes located on nested
topological spheres surrounding the galaxy.  The catastrophe structure of
the inner caustics depends on the angular momentum distribution of the
infalling dark matter particles.  If that angular momentum distribution
is dominated by net overall rotation, implying 
$\vec{\nabla} \times \vec{v} \neq 0$, the inner caustics are a set of 
`tricusp rings'.  A tricusp ring is a closed tube whose cross section 
is a section of the {\it elliptic umbilic} ($D_{-4}$) catastrophe 
\cite{crdm,crsing,inner}.  The rings are located in the plane of the 
galaxy.  In the self-similar infall model \cite{FGB}, generalized to 
include the effect of angular momentum \cite{STW,MWhalo}, the caustic 
ring radii $a_n$ ($n =1,2,3 ..$) are predicted to obey the law 
\cite{crdm,MWhalo}
\begin{equation}
a_n \simeq {40 {\rm kpc} \over n}~
\left({v_{\rm rot} \over 220 {\rm km/s}}\right)
\left({j_{\rm max} \over 0.18}\right)~~~\ , 
\label{crr}
\end{equation}
where $v_{\rm rot}$ is the rotation velocity of the galaxy
and $j_{\rm max}$ is a parameter characterizing the amount
of angular momentum that the dark matter particles carry.

Observational evidence for caustic rings at the radii $a_n$
predicted by Eq. (\ref{crr}) was found in the rotation curves 
of external galaxies \cite{Kinn}, the rotation curve of our 
own galaxy \cite{milky}, and an IRAS map of the Galactic plane 
in the direction of the nearest caustic ring ($n=5$)\cite{milky}.
A summary of the evidence can be found in ref. \cite{MWhalo}. The 
evidence implies that the distribution of $j_{\rm max}$ values 
over nearby spiral galaxies, including the Milky Way, is peaked 
at $j_{\rm max} \simeq 0.18$.  It also implies that we on Earth 
are close to a cusp in the nearest caustic ring of dark matter.  
As a result, the dark matter velocity distribution on Earth is 
dominated by a single flow, of known velocity vector.  That 
single flow, called the ``Big Flow" has density of order 
1 GeV/cc, which is two or three times larger than the 
commonly cited estimates of the {\it total} local dark 
matter density. 

Finally, the evidence for caustic rings of dark matter halos 
implies that the dark matter particles fall in with net overall 
rotation, and hence that their velocity field has non-zero curl: 
$\vec{\nabla} \times \vec{v} \neq 0$.  If their velocity field 
were irrotational ($\vec{\nabla}\times\vec{v} = 0$), the inner 
caustics would have a tent-like structure \cite{inner} which is 
quite distinct from that of the caustic rings for which evidence 
was found.

This raises a puzzle.  Indeed if the dark matter is cold and 
collisionless, as is the case for weakly interacting massive 
particles (WIMPs) and was thought to be the case for axions, 
the velocity field remains irrotational at all times because 
it is the outcome of gravitational forces proportional to the 
gradient of the Newtonian potential \cite{inner}.  (General 
relativistic effects allow the creation of rotational velocity 
fields but are subdominant because the velocities involved are 
much less than the speed of light.)  Thus, if the dark matter 
is cold and collisionless, one expects the inner caustics of 
galactic halos to be the tent-like structures of the 
$\vec{\nabla}\times\vec{v} = 0$ case, instead of the rings 
for which evidence was found.  This puzzle has bothered me for 
a number of years, but Qiaoli Yang and I may now have found a 
solution to it \cite{CABEC}.  As explained below, cold dark 
matter axions form a Bose-Einstein condensate.  As a result, 
their properties differ from those of ordinary CDM. 

\section{Bose-Einstein condensation of dark matter axions}~\footnote{All 
the material in this section is taken from ref. \cite{CABEC}.} 

The number density of cold axions implied by Eq.~(\ref{nat1}) is 
\begin{equation}
n(t) \sim {4 \cdot 10^{47} \over {\rm cm}^3}~
\left({f \over 10^{12}~{\rm GeV}}\right)^{5 \over 3}
\left({a(t_1) \over a(t)}\right)^3
\label{numden}
\end{equation}
where $a(t)$ is the cosmological scale factor.  Because the axion
momenta are of order ${1 \over t_1}$ at time $t_1$ and vary with
time as $a(t)^{-1}$, the velocity dispersion of cold axions is
\begin{equation}
\delta v (t) \sim {1 \over m t_1}~{a(t_1) \over a(t)}
\label{veldis}
\end{equation}
{\it if} each axion remains in whatever state it is in, i.e. if axion
interactions are negligible.  Let us refer to this case as the limit of
decoupled cold axions.  If decoupled, the average state occupation number
of cold axions is
\begin{equation}
{\cal N} \sim~ n~{(2 \pi)^3 \over {4 \pi \over 3} (m \delta v)^3}
\sim 10^{61}~\left({f \over 10^{12}~{\rm GeV}}\right)^{8 \over 3}~~\ .
\label{occnum}
\end{equation}
Clearly, the effective temperature of cold axions is much smaller than
the critical temperature
\begin{equation}
T_{\rm c} = \left({\pi^2 n \over \zeta(3)}\right)^{1 \over 3}
\simeq 300~{\rm GeV}~\left({f \over 10^{12}~{\rm GeV}}\right)^{5 \over 9}~
{a(t_1) \over a(t)}
\label{Tc}
\end{equation}
for BEC.  Axion number violating processes, such as their decay to two
photons, occur only on time scales vastly longer than the age of the
universe.  The only condition for axion BEC that is not clearly
satisfied is thermal equilibrium.

Axions are in thermal equilibrium if their relaxation rate $\Gamma$ is
large compared to the Hubble expansion rate $H(t) = {1 \over 2t}$.  At
low phase space densities, the relaxation rate is of order the particle
interaction rate $\Gamma_s = n \sigma \delta v$ where $\sigma$ is the
scattering cross-section.  Axions have self interactions described by the 
action density ${\cal L}_{\rm self} = + {1 \over 4\!} \lambda \varphi^4$
where $\lambda \simeq 0.35 ({m \over f})^2$.  The cross-section for
$\varphi + \varphi \rightarrow \varphi + \varphi$ scattering due to
axion self interaction is {\it in vacuum}
\begin{equation}
\sigma_0 = {1 \over 64 \pi} {\lambda^2 \over m^2} \simeq
1.5 \cdot 10^{-105} {\rm cm}^2 \left({m \over 10^{-5}~{\rm eV}}\right)^6~~~\ .
\label{xs0}
\end{equation}
If one substitutes $\sigma_0$ for $\sigma$, $\Gamma_s$ is found much
smaller than the Hubble rate, by many orders of magnitude.  However,
in the cold axion fluid background, the scattering rate is enhanced by
the average quantum state occupation number of both final state axions,
$\sigma \sim \sigma_0 {\cal N}^2$, because energy conservation forces 
the final state axions to be in highly occupied states if the initial
axions are in highly occupied states.  In that case, the relaxation rate
is multiplied by {\it one} factor of ${\cal N}$ \cite{ST}
\begin{equation}
\Gamma \sim n~\sigma_0~\delta v~{\cal N}~~~\ .
\label{rate}
\end{equation}
Combining Eqs.~(\ref{numden}-\ref{occnum},\ref{xs0}), one finds
$\Gamma(t_1)/H(t_1) \sim {\cal O}(1)$, suggesting that cold axions
thermalize at time $t_1$ through their self interactions, but only
barely so.

It may seem surprising that the huge and tiny factors on the RHS of
Eq.~(\ref{rate}) cancel each other.  In fact the cancellation is not
an accident.  Consider a generic axion-like particle (ALP) whose mass
$m$ and decay constant $f$ are unrelated to each other.  Its self
interaction coupling strength $\lambda \sim {m^2 \over f^2}$.  Cold
ALPs appear at a time $t_1 \sim {1 \over m}$ with number density
$n(t_1) \sim f^2 m$, and velocity dispersion $\delta v (t_1) \sim 1$.
Substituting these estimates in Eqs. (\ref{occnum}), (\ref{xs0}) and
(\ref{rate}), one finds that the thermalization rate is of order the
Hubble rate at $t_1$, for all $f$ and $m$.

A critical aspect of axion BEC phenomenology is whether the BEC
continues to thermalize after it has formed.  Axion BEC means
that (almost) all axions go to one state.  However, only if the
BEC continually rethermalizes does the axion state track the
lowest energy state.

The particle kinetic equations that yield Eq.~(\ref{rate}) are
valid only when the energy dispersion ${1 \over 2} m (\delta v)^2$
is larger than the thermalization rate \cite{ST}.  After $t_1$ this   
condition is no longer satisfied.  One enters then a regime where
the relaxation rate due to self interactions is of order
\begin{equation}
\Gamma_\lambda  \sim \lambda~n~m^{-2}~~\ .
\label{rate2}
\end{equation}
$\Gamma_\lambda(t)/H(t)$ is of order one at time $t_1$ but
decreases as $t~a(t)^{-3}$ afterwards.  Hence, self interactions
are insufficient to cause axion BEC to rethermalize after $t_1$   
even if they cause axion BEC at $t_1$.  However gravitational
interactions, which are long range, come in to play.  The
relaxation rate due to gravitational interactions is of order
\begin{equation}
\Gamma_{\rm g} \sim G~n~m^2~\ell^2
\label{rate3}
\end{equation}
where $\ell \sim (m \delta v)^{-1}$ is the correlation length.
$\Gamma_{\rm g}(t)/H(t)$ is of order
$4 \cdot 10^{-8}(f/10^{12}~{\rm GeV})^{2 \over 3}$
at time $t_1$ but grows as $t a^{-1}(t) \propto a(t)$.  Thus
gravitational interactions cause the axions to thermalize and
form a BEC when the photon temperature is of order
100 eV~$(f/10^{12}~{\rm GeV})^{1 \over 2}$.

The process of axion Bose-Einstein condensation is constrained by
causality.  Thus one expects overlapping condensate patches with 
typical size of order the horizon.  As time goes on, say from $t$ 
to $2t$, the axions in $t$-size condensate patches rethermalize 
into $2t$-size patches.  The correlation length is then of order 
the horizon at all times, implying $\delta v \sim {1 \over m t}$ 
instead of  Eq.~(\ref{veldis}), and 
$\Gamma_{\rm g}/H \propto t^3 a^{-3}(t)$  
after the BEC has formed.  Therefore gravitational interactions
rethermalize the axion BEC on ever shorter time scales compared  
to the age of the universe.

The axion field may be expanded in modes
labeled $\vec\alpha$:
\begin{equation}
\varphi(x) = \sum_{\vec\alpha}~[a_{\vec\alpha}~\Phi_{\vec\alpha}(x) 
~+~a_{\vec\alpha}^\dagger~\Phi_{\vec\alpha}^\star]
\label{modex}
\end{equation}
where the $\Phi_{\vec\alpha}(x)$ are the positive frequency c-number  
solutions of the Heisenberg equation of motion for the axion field
\begin{equation}
D^\mu D_\mu \varphi(x) = g^{\mu\nu}[\partial_\mu \partial_\nu -
\Gamma_{\mu\nu}^\lambda \partial_\lambda] \varphi(x) = m^2 \varphi(x)~~~\ ,
\label{eom}
\end{equation}
and the $a_{\vec\alpha}$ and $a_{\vec\alpha}^\dagger$ are creation
and annihilation operators satisfying canonical commutation relations.
We neglect the self-interaction term  which would otherwise appear on the 
RHS of Eq.~(\ref{eom}), because it is of order ${\rho \over f^2} \varphi$, 
where $\rho$ is the axion density, and hence smaller by the factor
$\left({a(t_1) \over a(t)}\right)^3 {t \over t_1}$ than the relevant
terms (of order ${m \over t} \varphi$) in that equation.  BEC means
that all cold axions, except for a small fraction, go to a single 
state which we label $\vec\alpha = 0$.  The corresponding 
$\Phi_0(x)$ is the axion wavefunction.  In the spatially flat, 
homogeneous and isotropic Robertson-Walker space-time,
\begin{equation}
\Phi_0 = {A \over a(t)^{3 \over 2}}~e^{-i m t}
\label{RW}
\end{equation}
where $A$ is a constant.  The state of the axion field is
$|N> = (1/\sqrt{N!})~(a_0^\dagger)^N |0>$ where $|0>$ is the
empty state, defined by $a_{\vec\alpha}~|0>$ = 0 for all
$\vec\alpha$, and $N$ is the number of axions.

To compare axion BEC with CDM, let us divide the observations into 
three arenas: 1) the behaviour of density perturbations on the scale 
of the horizon, 2) their behaviour during the linear regime of evolution 
within the horizon, and 3) their behaviour during the non-linear regime.  
CDM provides a very successful description in arena 2.  However, axion 
BEC and CDM are indistinguishable in arena 2 on all scales of observational 
interest \cite{CABEC,Hwang}.  In particular, the equation governing the 
evolution of axion BEC perturbations is 
\begin{equation}
\partial_t^2 \delta + 2 H \partial_t \delta
- \left(4 \pi G \rho_0 - {k^4 \over 4 m^2 a^4}\right) \delta = 0
\label{deneq}
\end{equation}
where $k$ is co-moving wavevector.  The last term in Eq.~(\ref{deneq}) 
is absent for CDM.  Eq.~(\ref{deneq}) implies that the axion BEC has 
Jeans length
\begin{eqnarray}
k_{\rm J}^{-1} &=& (16 \pi G \rho m^2)^{-{1 \over 4}}\nonumber\\
&=& 1.02 \cdot 10^{14}~{\rm cm}
\left({10^{-5}~{\rm eV} \over m}\right)^{1 \over 2}
\left({10^{-29}~{\rm g/cm^3} \over \rho}\right)^{1 \over 4}~\ .
\label{Jeans}
\end{eqnarray}
However, the Jeans length is small compared to the smallest scales 
($\sim$ 100 kpc) for which we have observations on the behavior of 
density perturbations in the linear regime.  

In the non-linear regime of structure formation (arena 3) and
in the absence of rethermalization, axion BEC and CDM again 
differ only on length scales smaller than de Broglie wavelength.
This follows from the WKB approximation and has also been shown 
by numerical simulation \cite{WK}.  Since the axion de Broglie 
wavelength (of order 10 meters in galactic halos) is negligbly 
small compared to all length scales of observational interest, 
we again find that axion BEC and CDM are indistinguishable when 
there is no rethermalization of the BEC.

However, it was found above that gravitational interactions do rethermalize 
the axion BEC continually so that the axion state tracks the lowest energy
state.  This is relevant to the angular momentum distribution of dark
matter axions in galactic halos.  The angular momentum of galaxies is 
caused by the gravitational torque of nearby galaxies early on when 
protogalaxies are still close to one another \cite{Peebles}.  As was 
mentioned in Section III, CDM presents us with a puzzle.  The velocity 
field of ordinary cold dark matter, such as WIMPs, remains irrotational 
whereas the evidence for caustic rings of dark matter implies that the 
dark matter falls in with net overall rotation.  The puzzle is solved 
if the dark matter is an axion BEC which rethermalizes while tidal 
torque is applied to it.  Indeed, the lowest energy state for given 
total angular momentum is one in which each particle carries an equal
amount of angular momentum.  In that case there is net overall
rotation.  $\vec{\nabla} \times \vec{v} \neq 0$ is accomodated
in the BEC through the appearance of vortices.  The phenomenon
is observed in quantum liquids and well understood \cite{Pethik}.

Finally let's consider the behaviour of density perturbations as they 
enter the horizon (arena 1).  Here too axion BEC differs from CDM.  The 
CDM perturbations evolve linearly at all times.  The axion BEC perturbations
do not evolve linearly when they enter the horizon because the condensates
which prevailed in neighboring horizon volumes rearrange themselves, through
their gravitational interactions, into a new condensate for the expanded
horizon volume.  This produces local correlations between modes of different
wavevector since the perturbation of wavevector $\vec{k}$, upon entering
the horizon, is determined by the perturbations of wavevector say  
${1 \over 2}\vec{k}$ in its neighborhood.  We propose this as a  
mechanism for the alignment of CMBR anisotropy multipoles \cite{align}
through the integrated Sachs-Wolfe (ISW) effect. Unlike CDM, the ISW
effect is large in axion BEC because the Newtonian potential $\psi$
changes entirely after entering the horizon in response to the
rearrangement of the axion BEC.

\section{References}



\end{document}